\documentstyle[12pt]{article}

\def\beq{\begin{equation}}

\def\eeq{\end{equation}}

\def\bea{\begin{eqnarray}}

\def\eea{\end{eqnarray}}

\def\bq{\begin{quote}}

\def\eq{\end{quote}}

\def\nnb{\nonumber}

\def\ga{\left(}

\def\dr{\right)}

\def\rar{\rightarrow}

\def\nnb{\nonumber}

\def\la{\langle}

\def\ra{\rangle}

\def\nin{\noindent}
\def\ba{\begin{array}}
\def\ea{\end{array}}

\begin{document}

\topmargin -2.5cm

\oddsidemargin -.5cm

\evensidemargin -1.0cm

\pagestyle{empty}

\begin{flushright}
PM 96/26\\
\end{flushright}

\vspace*{0.5cm}

\begin{center}
\section*{Heavy quarkonia mass-splittings in QCD:\\
test of the $1/m$-expansion and\\
estimates of $\la\alpha_s G^2\ra$ and $\alpha_s$}

\vspace*{0.5cm}
{\bf S. Narison}\\
Laboratoire de Physique Math\'ematique,\\
Universit\'e de Montpellier II, Place Eug\`ene Bataillon,\\ 
34095 - Montpellier Cedex 05, France.

\vspace*{1.5cm}
{\bf Abstract} \\ \end{center}

\nin
I present a more refined analysis of the mass-splittings
between the different
heavy quarkonia states, using {\it new double ratios} 
of exponential moments of different
two-point functions. Then, I test the validity of
the $1/m$-expansion, extract
$\alpha_s(M_Z)=0.127\pm 0.011$ from 
$M_{\chi_c(P^1_1)}-M_{\chi_c(P^3_1)}$, and provide a new estimate  
of the gluon condensate from $M_\psi
-M_{\eta_c}$ and $M_{\chi_b}-M_\Upsilon$, which combined with
the recent estimate from the $\tau$-like decay sum rules in
$e^+e^-\rar I=1$ hadrons data, leads to the {\it update average value}
$\la\alpha_s G^2\ra = (7.1\pm 0.9)\times 10^{-2}$ GeV$^4$ from the
light and heavy quark systems. 
I also 
find $M_{\Upsilon}-M_{\eta_b}\approx 63^{-29}_{+51}$ MeV implying the
possible observation of the $\eta_b$ in the $\Upsilon$-radiative decay.

\vspace{1.5cm}

\nin
{\it Talk given at QCD 96-Montpellier, 4-12th July 1996 
and at the 28th ICHEP96-Varsaw, 25-31th July 1996,
and based on the paper hep-ph/9512348 (to appear in
Phys. Lett.B386, October 1996), which will be referred as SN.}

\vspace{1.5cm}
\begin{flushleft}
PM 96/26\\
September 1996
\end{flushleft}
\vfill\eject

\setcounter{page}{1}

 \pagestyle{plain}
\section{The double ratio of moments} \par

\nin
QCD spectral sum rule (QSSR) \`a la SVZ \cite{SVZ} 
(for a recent
review, see e.g. \cite{SNB}) has shown since
15 years, its impressive ability
for describing the complex phenomena of hadronic
physics with the few universal ``fundamental" parameters
 of the QCD
Lagrangian
(QCD coupling $\alpha_s$, quark masses
and  vacuum condensates built from the quarks
and/or gluon fields). 
In the example of the two-point correlator:
\beq
\Pi_Q(q^2) \equiv i \int d^4x ~e^{iqx} \
\la 0\vert {\cal T}
J_Q(x)
\ga J_Q(o)\dr ^\dagger \vert 0 \ra ,
\eeq
 associated to the generic hadronic current:
$J_Q(x) \equiv \bar Q \Gamma Q (x)$ of the heavy $Q$-quark 
($\Gamma$ is
a Dirac matrix which specifies the hadron quantum numbers),
the SVZ-expansion reads:
\bea
\Pi_Q (q^2)
&\simeq& \sum_{D=0,2,...}\sum_{dim O=D}
\frac{ C^{(J)}(q^2,\mu)\la O(\mu)\ra}
{\ga M^2_Q-q^2 \dr^{D/2}},\nnb\\ 
\eea
where $\mu$ is an arbitrary scale that separates the long- 
and
short-distance dynamics; $C^{(J)}$ are the Wilson 
coefficients 
calculable
in perturbative QCD by means of Feynman diagrams 
techniques;
$\la O \ra$ are
the non-perturbative condensates of dimension $D$ built
 from the quarks or/and gluon
fields ($D=0$
corresponds to the case of the na\"\i ve perturbative 
contribution). Owing to gauge invariance, the lowest
dimension condensates that can be formed are the $D=4$
light quark $m_q \la\bar \psi \psi \ra$ and gluon $\la
\alpha_s 
G^2 \ra$
ones, where the former is fixed by the pion PCAC relation, 
whilst 
the latter
is known to be $(0.07\pm 0.01)$ GeV$^4$ from more recent 
analysis
of the light \cite{SNL} quark systems.
The
validity of the SVZ-expansion can be understood, using
renormalon techniques (absorption of the IR renormalon 
ambiguity into the definitions
of the condensates and absence of
some extra $1/q^2$-terms not included 
in the OPE)
\cite{MUELLER,BENEKE} and/or by building  
renormalization-invariant
combinations of the condensates (appendix of \cite{PICH} 
and 
references
therein). The SVZ expansion is phenomenologically 
confirmed from (among other applications)
the unexpected
accurate determination of the QCD coupling $\alpha_s$ 
\cite{PICH}--\cite{DUFLOT} and 
from
a measurement of the condensates \cite{DUFLOT}
from semi-inclusive $\tau$-decays.
The previous QCD information is transmitted to the data 
through 
the spectral function Im$\Pi_Q(t)$
via the K\"allen--Lehmann dispersion relation ($global~
duality)$
 obeyed by the hadronic correlators,
which can be improved from the uses of different versions
of the sum rules \cite{SVZ},\cite{SNB},\cite{RRY}--\cite{SNR}. 
In this paper, we shall use the simple duality ansatz 
parametrization: $
``{one~narrow~resonance}"+
 ``{QCD~ continuum}"$, from a threshold $t_c$, which
gives a good description of the spectral integral in the
sum rule analysis, as has been
tested successfully in the light-quark channel 
from the 
$e^+e^- \rar$
$I=1$ hadron data and in the heavy-quark ones from the
$e^+e^- \rar \psi$ or $\Upsilon$ data. 
We shall work with the relativistic
version of the Laplace or exponential sum rules: 
\cite{SVZ,BELL,BERT,SNR}:
\bea
{\cal L_H}(\sigma,m^2)
&\equiv& \int_{4m^2}^{\infty} {dt}~\mbox{exp}(-t\sigma)
~\frac{1}{\pi}~\mbox{Im} \Pi_Q(t)
,\nnb\\
{\cal R}_H(\sigma)&\equiv& -\frac{d}{d\sigma}
\log{{\cal L}_H(\sigma,m^2)}
,
\eea
 where the 
QCD expression
known to order
$\alpha_s$ is given (without expanding in $1/m$)
in terms of the pole mass
$m(p^2=m^2)$,
\footnote{For consistency, we shall work with 
the
too-loop order $\alpha_s$ expression
of the pole mass \cite{SNM}.} and contains
the gluon condensate $\la \alpha_s G^2\ra$ correction
to leading order. To this order, the gluon condensate 
is well-defined as the ambiguity
only comes from higher order terms in $\alpha_s$, 
which have, however,
a smaller numerical effect than the one from the 
error of the
phenomenological estimate of the condensate.
 $\sigma \equiv \tau\equiv 1/M^2$ 
(notations used in the literature) is the
exponential Laplace sum rule variable; $H$ specifies the 
hadronic channel studied. 
In principle, the pair $(\sigma,t_c)$ is free 
external
parameters in the analysis, so that the optimal result 
should be
insensitive to their variations. {\it Stability criteria}, 
which 
are equivalent
to the variational method, state that the best results 
should
be obtained at the minimas or at the inflexion points in 
$\sigma$,
while stability in $t_c$ is useful to control the 
sensitivity of 
the
result in the changes of $t_c$-values. These stability 
criteria are satisfied in the heavy quark channels studied
here, as the continuum effect is negligible
and does not exceed 1\% of the ground
state contribution \cite{SNB,BERT}, such that 
at the minimum in $\sigma$, one expects
 to a good approximation:
\beq
\mbox{min}_{\sigma}{\cal R}(\sigma)\simeq M^2_H.
\eeq 
Moreover,
one can  $a ~posteriori$ check that, at the stability
point, where we have an equilibrium between the continuum 
and the
non-perturbative
contributions, which are both small,
the OPE is still convergent and validates the SVZ-expansion. 
The previous approximation can be improved
by working with the double ratio of moments 
\footnote{This method has also been used 
in \cite{SNBOT} for studying the mass
splittings of the heavy-light quark systems.} 
:
\bea
{\cal R}_{HH'}(x)\equiv \frac{{\cal R}_H}{{\cal R}_{H'}}
\simeq \frac{M^2_H}{M^2_{H'}},
\eea
provided that each ratio of moments stabilizes at about 
the
same value of $\sigma$. In this case, there is a
cancellation of the different leading terms like the 
heavy quark mass
(and its ambiguous definition used in some previous literatures), 
the negligible continuum 
effect (which is already small in the ratio of
moments), and each leading QCD corrections. \\ 
\section{Test of the $1/m$-expansion}

\nin
For this purpose, we use the complete {\it horrible}
results expressed in terms of the
pole mass to order $\alpha_s$ given by \cite{BERT} and checked by 
various authors
\cite{SNB}, which we expand in series of $1/m$
with the help of the Mathematica 
program. By comparing the complete and truncated series in $1/m$,
one can notice that, at the $c$ and $b$ mass scales, the convergence of 
the $1/m$-expansion is quite bad due to the increases of the numerical 
coefficients with the power of $1/m$ and to the alternate signs 
of the $1/m$ series. This feature invalidates the analysis in Ref. \cite{DOMI}.
\section{Balmer-mass formula}

\nin
The Balmer formula derived from a
non-relativistic approach ($m\rar \infty$)
of the Schr\"odinger levels 
reads for the $S^3_1$ vector meson
\cite{LEUT} (see also \cite{DOSCH},\cite{YND}):
\beq
M_{S^3_1}^{static}\simeq 2m
\Bigg{[}1-\frac{2}{9}\alpha_s^2+
0.23\frac{\pi}
{(m\alpha_s)^4}\la\alpha_s G^2\ra\Bigg{]}.
\eeq

\nin
It is instructive to compare this result with the mass
formula obtained from the ratio of moments 
within the $1/m$-expansion. Using the different QCD
corrections, one obtains the mass formula
at the minimum in $\sigma$ of ${\cal R}$:
\beq
M_{\Upsilon}\simeq 2m_b\ga 1+\frac{\pi}{27}\alpha_s^2\dr
M^{static}_{S^3_1},
\eeq
at the value:
\beq
\sqrt{\sigma_{coul}}\simeq \frac{9}{4m\alpha_s\sqrt{\pi}}\simeq
0.85 ~\mbox{GeV}^{-1}~.
\eeq
One can recover by identification in the 
{\it static limit}
($m_b\rar \infty$) the previous Balmer formula, 
but a new extra $\alpha_s^2$ correction 
due to the 
$v^2$ (finite mass) terms in the
free part appears here
 (for some derivations of the relativistic correction
in the potential approach see \cite{BYERS}, \cite{ZAL2}), 
and tends to reduce 
the coulombic interactions. On the other hand, at the 
$b$-quark
 mass scale, the dominance of the gluon condensate contribution 
indicates that the $b$-quark is not enough heavy
for this system to be coulombic rendering
the non-relativistic potential approach to be a
crude approximation at this scale. 
The extension of this comparison to the cases of the $S^3_1$-$S^1_0$
hyperfine and $S$-$P$ splittings
is not very conclusive, as in the former, one needs
an evaluation of the correlator at the next-next-to-leading 
order for a better control of the $\alpha^2_s$ terms, 
while, in the latter 
there is a discrepancy for the coefficients of the gluon 
condensate 
in the two approaches, which may reflect
the difficulty of Bell-Bertlmann \cite{BELL,BERT} to find a 
bridge between the field theory \`a la SVZ
(flavour-dependent confining potential) 
and the potential models (flavour-independence). 

\section{Leptonic width and wave 
function }

\nin
Using the sum rule ${\cal L}_H$ and saturating it by 
the vector $S^3_1$ state, we obtain, to a good approximation,
 the sum rule:
\bea
&&M_V\Gamma_{V\rar e^+e^-}\simeq {(\alpha e_Q)^2}
\frac{e^{2\delta m M_V \sigma}}{72\sqrt{\pi}}
\frac{\sigma^{-3/2}}{m}\nnb\\
&&\Bigg{[} 1+\frac{8}{3}\sqrt{\pi\sigma}m\alpha_s
-\frac{4\pi}{9}\la \alpha_s G^2\ra m\sigma^{5/2}
\Bigg{]}~, 
\eea
where $e_Q$ is the quark charge in  units of e; $\delta 
m \equiv
M_V-2m$ is the meson-quark mass gap. In 
 the case of the $b$-quark, we use \cite{SNM} $\delta m \simeq 
0.26 $ GeV,
and the value of $\sigma_{min}$ given in Eq. (8). Then:
\beq
\Gamma_{\Upsilon(S^3_1)\rar e^+e^-}\simeq 1.2~\mbox{keV}~,
\eeq
in agreement with the data 1.3 keV. However, one should remark 
from Eq. (9), that the $\alpha_s$ correction 
is huge and needs an evaluation of the higher
order terms (the gluon condensate effect is negligible),
while the exponential factor effect is large, such that 
one can
{\it reciprocally} use the data on the width to fix either 
$\alpha_s$ or/and the quark mass \footnote{Larger value of the 
heavy quark
mass at the two-loop level
corresponding to a negative value of $\delta m$, would imply a 
much smaller
value of the leptonic width in disagreement with the data.}.
In the non-relativistic approach,
one can express the quarkonia
leptonic width in terms of its wave function $\Psi(0)_Q$,
from which, one can deduce:
\beq
16\pi|\Psi(0)|^2_Q \ga 1-4C_F\frac{\alpha_s}{\pi}\dr
\simeq 18.3~\mbox{GeV}^3~. 
\eeq
Using the expression of $\sigma_{coul}$, one can find 
that,
to leading order, the two approaches give 
a similar behaviour  for $\Psi(0)_Q$
in $\alpha_s$ and in $m$
 and about the same value of this quantity. 
However, one should notice that
in the present approach, the
QCD coupling $\alpha_s$ is evaluated at the scale $\sigma$
as dictated by the renormalization group equation obeyed
by the Laplace sum rule \cite{SNR} but $not$ at the resonance mass!
\section{ Gluon condensate
from $M_{\psi(S^3_1)}-M_{\eta_c(S^1_0)}$}

\nin
The value of $\sigma$, at which, the $S$-wave charmonium 
ratio of sum rules stabilize is: 
\cite{BERT}:
$
\sigma \simeq (0.9\pm 0.1) ~\mbox{GeV}^{-2}.
$
We use the range of the charm quark  
pole mass value $
m_c\simeq 1.2$-$1.5~\mbox{GeV}
$ to order $\alpha_s$
accuracy \cite{SNM} \footnote{For a recent 
review on 
the heavy quark masses, see e.g. \cite{ROD,PDG}.},
 and the double ratio of the vector 
$V(S^3_1)$ and the pseudoscalar $P(S^1_0)$ 
moments, which controls
the ratio of the meson mass squared. The exact 
expressions of the relativistic, Coulombic and
gluon condensate
lead respectively to corrections about 0.5, 2 and 7 $\%$ of 
the leading order one. One can understand from the 
approximate 
expressions in $1/m$ that the leading corrections
appearing in the ratio of moments cancel in
the double ratio, while the remaining
ones are relatively small. 
However, the expansion is not convergent for
the $\alpha_s$-term at the charm mass.
Using for 4 flavours \cite{SNM}: $
\alpha_s(\sigma)\simeq 0.48^{+ 0.17}_{-0.10}$,
and the experimental data \cite{PDG}:
${\cal R}^{exp}_{VP}= 1.082,$
one can deduce the value of the gluon condensate:
\beq
\la \alpha_s G^2\ra \simeq (0.10\pm 0.04)~\mbox{GeV}^4.
\eeq
We have estimated the error due to higher order effects 
by replacing the coefficient of $\alpha_s$ with the one
obtained from the effective Coulombic potential, which
tends to reduce the estimate to 0.07 GeV$^4$.
We have tested the convergence of the QCD series in 
$\sigma$,
by using the numerical estimate of the dimension-six
gluon condensate $g\la f_{abc}G^aG^bG^c\ra$ contributions
given in \cite{PARK}. This effect is about 0.1$\%$
of the zeroth order term and does not influence
the previous estimate in Eq. (12), which also indicates 
the
good convergence of the ratio of exponential moments 
already
at the charm mass scale in contrast with the $q^2=0$ 
moments
studied in Ref. \cite{SVZ,RADYU}. We also expect that in 
the double ratio of moments used here, the radiative 
corrections to the gluon condensate effects 
(available in the
literature \cite{BROAD})
are much smaller than in the
individual moments , such that they will give a 
negligible 
effect in the estimate of the gluon condensate.  
\section{Charmonium $P$-wave splittings}

\nin
The analysis of the different ratios of moments for the
$P$-wave charmonium shows 
\cite{BELL,BERT,PARK} that they are optimized for:
$
\sigma\simeq (0.6\pm 0.1)~\mbox{GeV}^{-2}.
$
The predictions for the scalar $P^3_0$ - axial $P^3_1$
and the tensor $P^3_2$-axial $P^3_1$ mass 
splittings, given in SN,
are satisfactory within 
our approximation.
\section{\bf $\alpha_s$ from the $P^1_1$ - $P^3_1$ 
axial mass splitting}

\nin
 The corresponding double ratio of moments has the nice 
feature 
to be independent of the gluon condensate to leading 
order in $\alpha_s$ and reads:
\beq
\frac{M_{P^1_1}^2}{M_{P^3_1}^2}\simeq 1+\alpha_s\Bigg{[}
\Delta_\alpha^{13}(\mbox{exact})=0.014^{-0.004}_{+0.008}\Bigg{]}~.
\eeq
The recent experimental value 3526.1 MeV of the $P^1_1$ mass denoted by
$h_c(1P)$ in the PDG compilation \cite{PDG} almost coincides with
the one of the center of mass energy,
as expected from the short range nature of the spin-spin force
\footnote{See e.g. \cite{MARTIN}.}. 
Using a na\"{\i}ve exponential
resummation of the higher order $\alpha_s$ terms, we deduce:
\bea 
\alpha_s(\sigma^{-1}\simeq
1.3~\mbox{GeV})\simeq 0.64^{+0.36}_{-0.18}\pm 0.02
\eea
which implies:
\bea
\alpha_s(M_Z)\simeq 0.127\pm 0.009 \pm 0.002, 
\eea
where
the error is much bigger than the one from LEP and $\tau$ decay data,
but its value is perfectly consistent with the latter. The theoretical error
is mainly due to the uncertainty in $\Delta_{\alpha}$, while a na\"{\i}ve
resummation of the higher order $\alpha_s$ terms leads the second error. 
Though inaccurate,
this value of $\alpha_s$ is interesting for
an alternative derivation of this fundamental quantity at lower energies, 
as it can
serve for testing its $q^2$-evolution until $M_Z$.
{\it Reciprocally}, using the value of $\alpha_s$ from LEP and 
$\tau$-decay
data as input, one can deduce the prediction of the center of 
mass (c.o.m) of the $P^3_J$ states given in Table 2 of SN. 
\section{$\Upsilon-\eta_b$ mass splitting}

\nin
For the bottomium, the analysis of the ratios of moments 
for the 
$S$ and $P$ waves shows that they are optimized at the 
same
value of $\sigma$, namely \cite{BERT}:
$
\sigma= (0.35\pm 0.05)~\mbox{GeV}^{-2},
$
which implies for 5 flavours: $
\alpha_s(\sigma)\simeq 0.32\pm 0.06$.
We shall use the conservative values of the two-loop
$b$-quark pole mass:
$m_b \simeq 4.2-4.7 ~\mbox{GeV}$. 
The splitting between the vector $\Upsilon(S^3_1)$ and the
pseudoscalar $\eta_b(S^1_0)$ can be done in a similar
way than the charmonium one.
One should also notice that, to this approximation, 
the gluon
condensate gives still the dominant effect at the $b$-mass 
scale (0.2$\%$ of the leading order) compared
to the one $.08\%$ from the $\alpha_s$-term. 
However, the $1/m$ series of the QCD $\alpha_s$
correction is badly convergent, showing that the static limit 
approximation can be quantitavely inaccurate in this channel. Therefore, 
one expects that the corresponding prediction of $(13^{-7}_{+10})$ MeV 
is a very crude estimate. In order to 
control the effect of the unknown higher order terms,
it is legitimate to introduce into the sum rule,
the coefficient of the 
Coulombic effect from the QCD potential \cite{YND}, which
leads to the ``improved" 
final estimate:
\beq
M_{\Upsilon}-M_{\eta_b}\approx \ga 63^{-29}_{+51}\dr~\mbox{MeV},
\eeq
implying the possible observation of the $\eta_b$ from the $\Upsilon$ 
radiative decay.
\section{Gluon condensate from $M_{\Upsilon}-M_{\chi_b}$}

\nin
As the $S$ and $P$ wave ratios of moments are optimized 
at the
same value of $\sigma$, we can compare directly, with a 
good
accuracy, the different $P$ states with 
the $\Upsilon~(S^3_1)$ one. As the coefficients of the 
$\alpha_s^2$ corrections, after inserting the expression
of $\sigma_{coul}$, 
are comparable with the one from the Coulombic
potential, we expect that the prediction of this 
splitting
is more accurate than in the case of the hyperfine.
The different double ratios of
moments leads to the
predictions of the $\chi_b$ states $P^3_0,~
P^3_1,~P^3_2$ given in Table 2 of SN in good agreement 
with the data, if one uses the value of $\alpha_s$ \cite{BETHKE}
and of the gluon condensate obtained previously.
Reciprocally, one can use the data for
re-extracting {\it independently} the value of the gluon
condensate. As usually observed in the literature, the 
prediction
is more accurate for the center of mass energy, 
than for the individual
mass. The corresponding numerical sum rule for:
\bea
\ga {M_{\chi_b}^{c.o.m}-M_{\Upsilon}}\dr/{M_{\Upsilon}}
\eea
is due to 
$+\ga 1.53^{+0.26}_{-0.42}\dr \times 10^{-2}$ of the relativstic
effects, $+\ga 1.20^{+0.1}_{-0.2}\dr \times 10^{-2}$ of the Coulombic
and $+(0.28^{+0.08}_{-0.06} )\mbox{GeV}^{-4}
\la\alpha_s G^2\ra~$ of the gluon condensate ones.
It leads to:
\beq
\la \alpha_s G^2\ra \simeq (6.9\pm 2.5)\times 
10^{-2}~\mbox{GeV}^4.
\eeq
We expect that this result is more reliable than the one 
obtained
from the $M_\psi-M_{\eta_c}$ as the latter can be more
affected by the non calculated next-next-to-leading 
perturbative
radiative corrections than the former. 
\section{Update average value of $\la\alpha_s G^2\ra~$}

\nin
Considering the
most recent estimate $(7\pm 1)\times 
10^{-2}~\mbox{GeV}^4$
from $e^+e^-\rar I=1$ hadrons data using
$\tau$-like decay \cite{SNL} as an update of the different
estimates from the light quark systems (see Table 2 of SN),
we can deduce from Eqs. (12) and (18),
the {\it update average} 
from a global fit of the light and heavy quark systems:
\beq
\la \alpha_s G^2\ra \simeq (7.1\pm 0.9)\times 
10^{-2}~\mbox{GeV}^4.
\eeq
This result confirms the
claim of Bell-Bertlmann
\cite{BELL,BERT} stating that
the SVZ value \cite{SVZ} has been underestimated by
about a factor 2 (see also \cite{FESR2,ZAL}).More accurate
measurements of this quantity than the already available
results from $\tau$-decay data \cite{DUFLOT} are needed for
testing the previous estimates from the
sum rules.
\section{Conclusions}

\nin
We have used {\it new double ratios} of exponential sum 
rules
for directly extracting  the mass-splittings of different 
heavy quarkonia states, the value of the gluon condensate
and of the QCD coupling $\alpha_s$.
Our numerical results are summarized in Eqs. (12), (18)
and (19) and in Table 2 of SN, 
where in the latter a comparison
with different estimates and experimental data is done.
We have also succeeded to derive the non-relativistic
Balmer formula from the sum rule, using a $1/m$-expansion,
where we have also included new
relativistic corrections due to finite value of the
quark mass. However, this expansion does not converge
at the $c$ and $b$ quark mass scale.
 

\begin{thebibliography}{999}
\bibitem{SVZ}
M.A. Shifman, A.I. Vainshtein and V.I.
Zakharov, {\it Nucl. Phys.} {\bf B147}
(1979) 385, 448.
\bibitem{SNB}
For a review, see e.g:
S. Narison, {\it QCD spectral sum rules,
Lecture Notes in Physics}, {\bf Vol. 26}
(1989) published by World Scientific.
\bibitem{SNL} S. Narison, 
hep-ph/9504334, {\it Phys. Lett.} {\bf B361} (1995) 121.
\bibitem{MUELLER}
A. Mueller, {\it QCD-20 Years Later Workshop, Aachen} 
(1992) 
and
references therein; F. David, {\it Montpellier
Workshop on Non-Perturbative Methods} ed. by World 
Scientific 
(1985);
M. Beneke and V.I. Zakharov,
{\it Phys. Rev. Lett.} {\bf 69 } (1992) 2472; G. 
Grunberg,
{\it QCD94}, Montpellier (1994).
\bibitem{BENEKE} P. Ball, M. Beneke and V. Braun, 
{\it Nucl. Phys.} {\bf B452} (1995) 563;
C.N. Lovett-Turner and C.J. Maxwell,
{\it Nucl. Phys.} {\bf B452} (1995) 188;
V. Zakharov, {\it QCD96 Montpellier}.
\bibitem{PICH}
E. Braaten, S. Narison and A. Pich,
{\it Nucl. Phys.} {\bf B373} (1992) 581.
\bibitem{ALFA}
F. Le Diberder and A. Pich,
{\it Phys. Lett.} {\bf B286} (1992) 147;  
D. Buskulic et al.,
{\it Phys. Lett.} {\bf B307} (1993) 209;  
A. Pich, {\it QCD94 Workshop}, Montpellier,
{\it Nucl. Phys. (Proc. Suppl)} {\bf B39} (1995);
S. Narison,
 {\it Tau94 Workshop}, Montreux,
{\it Nucl. Phys. (Proc. Suppl)} {\bf B40} (1995);
M. Girone and M. Neubert, hep-ph/9511392.
\bibitem{DUFLOT} L. Duflot, {\it Tau94 Workshop}, 
Montreux, {\it Nucl. Phys. (Proc. Suppl)} {\bf B40} 
(1995);
R. Stroynowski, ibid.
\bibitem{RRY}L.J. Reinders, H. Rubinstein and S. Yazaki,
{\it Phys. Rep.} {\bf 127} (1985) 1. 
\bibitem{BELL}
J.S. Bell and R.A. Bertlmann,{\it Nucl. Phys.} {\bf B177} (1981) 218;
{\bf B227} (1983) 435.
\bibitem{BERT} R.A. Bertlmann,
{\it Nucl. Phys.} {\bf B204} (1982) 387; 
{\it QCD90}, Montpellier,
{\it Nucl. Phys. (Proc. Suppl)} {\bf B23} (1991).
\bibitem{SNR}S. Narison and E. de Rafael, {\it Phys. Lett.} 
{\bf B103} (1981) 87.
\bibitem{SNM} S. Narison, {\it Phys. Lett.} {\bf B197} 
(1987) 405 {\bf B341} (1994) 73 and references therein.
\bibitem{DOMI}C.A. Dominguez and N. Paver,
 {\it Phys. Lett.} 
{\bf B293} (1992) 197; C.A. Dominguez, G.R. Gluckman
and N. Paver, {\it Phys. Lett.} 
{\bf B333} (1994) 184.
\bibitem{SNBOT} S. Narison, {\it Phys. Lett.} {\bf B210} 
(1988) 238. 
\bibitem{LEUT}H. Leutwyler, 
{\it Phys. Lett.} {\bf B98} (1981) 304.
\bibitem{DOSCH}H.G. Dosch and U. Marquard,
{\it Phys. Rev.} {\bf D35} (1987) 2238.
\bibitem{YND}F.J. Yndurain and S. Titard,
{\it Phys. Rev.} {\bf D} (1994) 231.
\bibitem{BETHKE}S. Bethke, {\it QCD96, Montpellier}; M. Schmelling,
{\it ICHEP, Varsaw} (1996).
\bibitem{PARK}J. Marrow and G. Shaw,
{\it Z. Phys.} {\bf C33} (1986) 237;
J. Marrow, J. Parker and G. Shaw,
{\it Z. Phys.} {\bf C37} (1987) 103.
\bibitem{BYERS}R. McLary and N. Byers,
{\it Phys. Rev.} {\bf D28} (1981) 1692.
\bibitem{ZAL2}For a review, 
see e.g. K. Zalewski, {\it QCD96, Montpellier}.
\bibitem{ROD}G. Rodrigo, Valencia preprint FTUV 95/30 
(1995), {\it Int. Work. Valencia} (1995).
\bibitem{PDG}PDG94 by L. Montanet et al. {\it Phys. Rev.} 
{\bf D50} (1994) 1173 and references therein.
\bibitem{RADYU}S.N. Nikolaev and A.V. Radyushkin,
{\it Phys. Lett.} {\bf B124} (1983) 243; 
{\it Nucl.Phys.} {\bf B213} (1983) 285.
\bibitem{BROAD}D.J. Broadhurst et al.
{\it Phys. Lett.} {\bf B329} (1994) 103. 
\bibitem{FESR2} R.A. Bertlmann, C.A. Dominguez, M. Loewe,
M. Perrottet and E. de Rafael,
{\it Z. Phys.} {\bf C39} (1988) 231.
\bibitem{ZAL}A. Zalewska and K. Zalewski, 
{\it Z. Phys.} {\bf C23} (1984) 233; K. Zalewski,
{\it Acta. Phys. Polonica} {\bf B16} (1985) 239 and
references therein.
\bibitem{SOLA}C.A. Dominguez and J. Sol\`a,
{\it Z. Phys.} {\bf C40} (1988) 63.
\bibitem{MARTIN} A. Martin, 21st ICHEP, Paris (1982)
{\it Journal de Physique, Colloque C3, supp} {\bf 12}, ed.
Editions de Physique, Les Ulysses; A. Martin, CERN-TH.6933/93,
Erice Lectures (1993); J.M. Richard,
{\it Phys. Rep.} {\bf 212} (1992) 1.
\bibitem{BUCH}W. Buchm\"uller, Erice Lectures (1984).
\bibitem{GIACO} G. Curci, A. Di Giacomo and G. Paffuti,
{\it Z. Phys.} {\bf C18} (1983) 135; M. Campostrini,
A. Di Giacomo and \~S. Olejn\'ik, Pisa preprint
IFUP-TH 2/86 (1986).
\end{thebibliography}
\end{document}